\documentclass[prl,twocolumn,showpacs,amsmath,amssymb,aps,superscriptaddress]
{revtex4-1}

\usepackage{graphicx}
\usepackage{dcolumn}
\usepackage{bm}
\usepackage{txfonts}
\usepackage{amsmath}
\usepackage{epsf}
\usepackage{epsfig}
\usepackage{amssymb}
\usepackage{color}
\usepackage{multirow}
\usepackage{pst-3d}
\usepackage{rotating}

\begin{document}

\title{Conditional Ramsey Spectroscopy with Synchronized Atoms}

\author{Minghui Xu and M. J. Holland}

\affiliation{JILA, National Institute of Standards and Technology and
  Department of Physics, University of Colorado, Boulder, Colorado
  80309-0440, USA}

\date{\today}%

\begin{abstract}
  We investigate Ramsey spectroscopy performed on a synchronized
  ensemble of two-level atoms. The synchronization is induced by the
  collective coupling of the atoms to a heavily damped mode of an
  optical cavity.  We show that, in principle, with this synchronized
  system it is possible to observe Ramsey fringes indefinitely, even
  in the presence of spontaneous emission and other sources of
  individual-atom dephasing. This could have important consequences
  for atomic clocks and a wide range of precision metrology
  applications.
\end{abstract}

\pacs{42.50.Pq, 03.65.Yz, 05.45.Xt, 06.30.Ft}

\maketitle

The precision currently achievable by atomic clocks is remarkable; for
example, the accuracy and instability of state-of-the-art optical
lattice clocks lies in the realm of
$10^{-18}$~\cite{Ludlow13,Jun14}. The pursuit of even more stability
is motivated by the potential benefit to a wide range of fields in the
physical and natural sciences, facilitating progress in diverse areas
such as; redefinition of the system of physical units in terms of
time~\cite{SI}, clock-based geodesy~\cite{geodesy}, gravitational wave
detection~\cite{gravitation}, and tests of fundamental physics and
cosmology~\cite{Bauch03,Derevianko13}. Atomic clock developments have
also enabled spin-off applications, including precision
measurements~\cite{Rosenband08}, quantum state
control~\cite{Leibfried03}, and investigations of quantum many-body
physics~\cite{Jun13,Ana14}.

\begin{figure}[b]
  \centerline{\includegraphics[width=0.7\linewidth, angle=0]{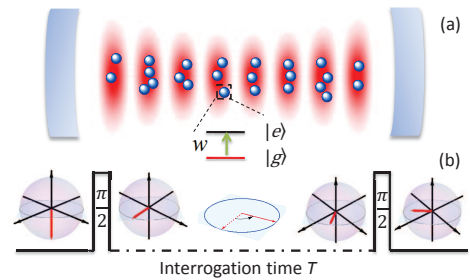}}
  \caption{\label{Fig1}(color online) (a) Conditional Ramsey
    spectroscopy where synchronized atoms are coupled collectively to
    a cavity and pumped individually with incoherent rate $w$. (b)
    Ramsey sequence showing preparation in state $|g\bigr>$
    (pseudospins pointing down to the south pole of the Bloch sphere),
    the $\pi/2$ $y$-axis rotation from the south pole to the equator,
    precession around the equator, and second $\pi/2$ $x$-axis
    rotation, after which the $z$-axis projection carries information
    about the cosine of the accumulated phase.}
\end{figure}

Atomic clocks typically operate using the method of Ramsey
Spectroscopy~(RS)~\cite{Ramsey56}. As shown in Fig.~\ref{Fig1}, RS
consists of three steps; (i) initial preparation of a coherent
superposition between two quantum states, (ii) accumulation of a phase
difference between the atoms and a local oscillator reference over an
interrogation time $T$, and (iii) mapping of the phase difference to a
population readout. Conventional RS is based on independent-atom
physics, with the role of a large number of atoms entering only
through improving the signal by statistical averaging. The performance
of RS is limited by the atomic coherence time, which causes decay of
the fringe visibility as a function of $T$. Due to this decay, an
optimal strategy is typically used that involves setting $T$ to be of
the order of the coherence time, and filling up the total measurement
interval $\tau$ by repeated RS cycles~\cite{cirac97}. This gives an
uncertainty in the frequency difference between the atoms and local
oscillator that scales as $1/(\sqrt{N\tau})$, with the $\sqrt{N}$
coming from the quantum projection noise at each readout. This scaling
$\tau^{-1/2}$ is much worse than the fundamental Fourier limit
$\tau^{-1}$.

There are two paths to improving on the standard limit for RS, apart
from simply increasing $N$. Firstly, the projection noise can be
reduced by preparing spin-squeezed
states~\cite{Wineland92,Ueda}. Pursuing this direction, there have
been numerous efforts to produce spin-squeezing in various physical
situations~\cite{h2,h1,Gross,pnas,Gross1,Riedel,mit1,cu}.  It is worth
pointing out that entangled states are often fragile and sensitive to
decoherence processes, which may limit their potential for providing
significant improvements to the
sensitivity~\cite{Cirac97,Lukin04}. Secondly, one can increase the
coherence time of atoms. One approach has been to increase the dephasing time
of magnetically and optically trapped atomic ensembles by spin self-rephasing induced
by the exchange interaction between two identical particles~\cite{Rosenbusch10,Rosenbusch11}. In recent lattice clock
experiments~\cite{Jun14}, the atomic dephasing time $T_2$ has been
pushed to $\sim$1s. Even if further technical improvements are made,
there is a fundamental upper limit to the atomic coherence time
provided by the lifetime,~$T_1$, of the long-lived excited clock
state~($\sim$160s for $^{87}$Sr)~\cite{Genes13}.

In this paper, we propose an approach to RS that is more robust
against decoherence. Our idea is to use atoms that resonantly exchange
photons with a heavily damped single-mode of an optical cavity during
the interrogation time of the RS sequence~[see
Fig.~\ref{Fig1}(a)]. Due to the cavity damping, it is necessary to
continuously replenish the energy by incoherently repumping the
atoms. One may have thought that this would simply give rise to
additional decoherence channels, on top of the usual $T_1$ and $T_2$
processes, and cause the RS fringe visibility to decay more
rapidly. This is not the case, since the cavity-mediated dissipative
coupling between atoms acts to synchronize their phases.  We show that
the coherence time of the synchronized ensemble does not depend on
individual-atom dephasing, as represented by $T_1$ and $T_2$. The
synchronized atoms instead undergo only a collective quantum phase
diffusion. However, the collective phase can be continuously monitored
by observing the cavity output field. Consequently, this system
provides a kind of conditional RS, conditioned on the cavity output, where fringes of high visibility
may be observed indefinitely.

The atom-cavity system during the interrogation time is described by
the Hamiltonian
\begin{equation}
\hat{H}=\frac{\hbar\Delta\nu} 2\sum_{j=1}^N\hat{\sigma}_j^{z}
+\frac{\hbar g} 2\sum_{j=1}^N(\hat{a}^\dagger\hat{\sigma}_j^-
+\hat{a}\hat{\sigma}_j^+),
\end{equation}
where $\Delta\nu$ is the frequency difference between the atoms and
local oscillator and $g$ is the coupling strength between a single
atom and the cavity mode. We introduce the bosonic annihilation and
creation operators, $\hat{a}$ and $\hat{a}^\dagger$, for cavity
photons, and the $j$-th atom Pauli operators, $\hat{\sigma}_j^z$ and
$\hat{\sigma}_j^-=(\hat{\sigma}_j^+)^\dagger$, for the pseudospins
representing the two-level system.  For simplicity, $g$ is assumed to
be identical for all atoms. In principle, this could be achieved by
trapping the atoms at the antinodes of the cavity mode by an optical
lattice. A less ideal spatial configuration only leads to a reduced
effective atom number, which has no impact on the basic conclusions of
this paper.

\begin{figure}[t]
  \centerline{\includegraphics[width=0.7\linewidth, angle=0]{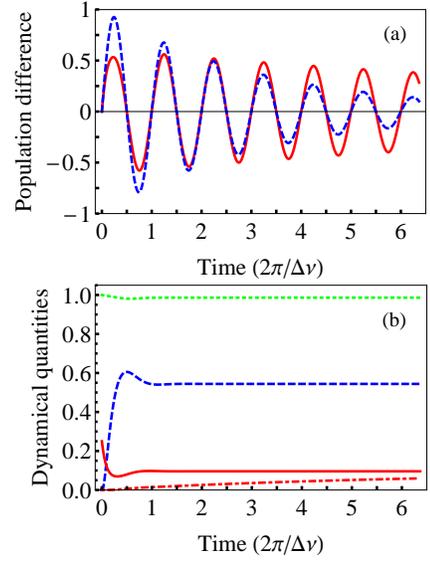}}
  \caption{\label{Fig2}(color online) Calculations of
    Eq.~(\ref{mastereq}) with $N=250$, $\Gamma_C=0.2/T_1$, $T_2=T_1$,
    and $w=N\Gamma_C/2$. (a) Ramsey fringes with synchronized
    atoms~(red solid line) versus $T$. Conventional Ramsey
    fringes~(blue dashed line) for the same $T_1$ and $T_2$. (b)
    During the interrogation time, the atomic inversion
    $\langle\hat{\sigma}_j^z\rangle$~(blue dashed line), spin-spin
    correlation $\langle\hat{\sigma}_j^+\hat{\sigma}_k^-\rangle$~(red
    solid line), $\langle\hat{\sigma}_j^+\hat{\sigma}_k^-\rangle$-
    $\langle\hat{\sigma}_j^+\rangle\langle\hat{\sigma}_k^-\rangle$~(red
    dotdashed line) and
    $\langle\hat{\sigma}_j^+\hat{\sigma}_k^z\rangle/
    (\langle\hat{\sigma}_j^+\rangle
    \langle\hat{\sigma}_k^z\rangle)$~(green dotted line).}
\end{figure}

In the presence of decoherence, the evolution is described by the
usual Born-Markov quantum master equation for the reduced atom-cavity
density matrix $\rho$,
\begin{equation}
\frac{d\rho}{dt}=
\frac{1}{i\hbar}[\hat{H},\rho]+\kappa\mathcal{L}[\hat{a}]\rho
+\sum_{j=1}^N\Bigl(w\mathcal{L}[\hat{\sigma}_{j}^+]
+\frac{1}{T_1}\mathcal{L}[\hat{\sigma}_{j}^-]
+\frac{1}{4T_2}
  \mathcal{L}[\hat{\sigma}_{j}^z]\Bigr)\rho
\end{equation}
where $\mathcal{L}[\hat{O}]\rho=(2\hat{O}\rho
\hat{O}^\dagger-\hat{O}^\dagger \hat{O}\rho-\rho \hat{O}^\dagger
\hat{O})/2$ denotes the Lindblad superoperator. The cavity decays with
rate $\kappa$ and the incoherent repumping is at rate
$w$. Conventional RS is recovered by setting $g=0$ and $w=0$, with the
result that the RS fringe visibility then decays exponentially with
the single-atom decoherence rate $\Gamma_S=(T_1^{-1}+T_2^{-1})/2$~[see
Fig.~\ref{Fig2}(a)].

We solve for the dynamics in an extreme regime of bad-cavity quantum
electrodynamics~\cite{Meiser09,Meiser101,Meiser102,Thompson12,thompson13},
where the vacuum Rabi splitting is much less than the cavity
linewidth, {\em i.e.}\/ $\sqrt{N}g\ll\kappa$. As a result, the cavity
is slaved to the atomic field and can be adiabatically
eliminated~\cite{Haake71}. The role of the cavity field then is to
simply provide a source for a dissipative collective coupling for the
atoms. The effective evolution is given by a quantum master equation
containing only atoms;
\begin{eqnarray}\label{mastereq}
  \frac{d\rho}{dt}&=&-\frac{i}{2}
  \Delta\nu\sum_{j=1}^N[\hat{\sigma}_j^{z},\rho]
  +\Gamma_C\mathcal{L}[\hat{J}^-]\rho
  \nonumber\\ &&{}
  +\sum_{j=1}^N\Bigl(w\mathcal{L}[\hat{\sigma}_{j}^+]
  +\frac{1}{T_1}\mathcal{L}[\hat{\sigma}_{j}^-]
  +\frac{1}{4T_2}
  \mathcal{L}[\hat{\sigma}_{j}^z]\Bigr)\rho,
\end{eqnarray}
where $\hat{J}^-=\sum_{j=1}^N\hat{\sigma}_j^-$ is the collective decay
operator  and $\Gamma_C=C/T_1$ is the collective decay rate, written
in terms of the cooperativity parameter of the cavity
$C$~\cite{kimble98}. The collective decay rate can be taken to be
small, {\em i.e.}\ $\Gamma_C\ll\Gamma_S$, because $C$ is a dimensionless
geometric cavity parameter that for real systems is typically much
less than 1.

It is extremely difficult to find numerical solutions to
Eq.~(\ref{mastereq}) for an appreciable number of atoms without
further approximation due to the exponential scaling, $4^N$, of the
dimensionality of the Liouvillian space. Fortunately, an underlying
{\em SU}(4) symmetry of the Liouvillian superoperators in
Eq.~(\ref{mastereq}) was developed recently, which reduces the
complexity of the problem to $N^3$~\cite{Xu13}. This enables us to
obtain numerical solutions up to a few hundred pseudospins.

Fig.~\ref{Fig2}(a) shows numerical calculations of RS fringes with
synchronized atoms. The solution of the quantum master equation
represents the ensemble average of many experimental trials. A
remarkable feature is that the fringe visibility decays much slower
than that of conventional RS under the same $T_1$ and $T_2$
decoherences, demonstrating the robustness to individual-atom
decoherence. When compared to conventional RS with independent atoms,
the principal difference here is that strong spin-spin correlations
between atoms $\langle\hat{\sigma}_j^+\hat{\sigma}_k^-\rangle$~($j\ne
k$) develop due to the dissipative coupling, as shown in
Fig.~\ref{Fig2}(b).  This feature is a characteristic of
phase-locking~\cite{Meiser09,Lee14}.  After a brief initial transient
evolution, the fringe fits well to an exponentially decaying $sine$
function, {\it i.e.}, $Ae^{-\lambda t}\sin\Delta\nu t$, where
$\lambda$ is the decay rate of the fringe visibility and $A$ is an
amplitude (we derive this behavior later.)

\begin{figure}[b]
  \centerline{\includegraphics[width=0.7\linewidth, angle=0]{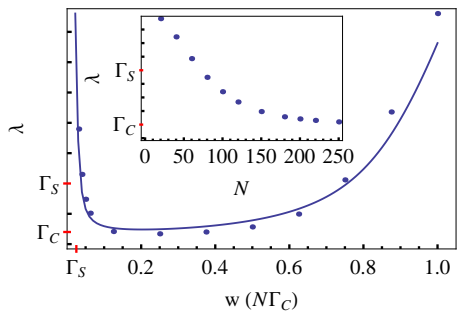}}
  \caption{\label{Fig3}(color online) The decay rate of the visibility
    of Ramsey fringes at $\Gamma_C=0.2/T_1$ and $T_2=T_1$ as a
    function of repumping for $N=200$ and as a function of $N$ for
    $w=N\Gamma_C/2$~(Inset). The dots are numerical solutions of
    Eq.~(\ref{mastereq}), and the solid blue line is the semiclassical
    approximation for comparison.}
\end{figure}

Intuitively, one may expect that in order to effectively phase-lock
the atoms, it should be necessary for the dissipative coupling that
provides rephasing to dominate over the `random-walk' due to quantum
noises that destroy phase correlations.  Because of the all-to-all
nature of the interaction of atoms through the cavity mode, the
dissipative coupling strength scales with $N$ and is given by
$N\Gamma_C/2$~\cite{Xu132}. We show the effect of this in the inset of
Fig.~\ref{Fig3}. For small atom number, the individual quantum noises
dominate over the rephasing, and the fringe envelope decays more
rapidly than in conventional RS, {\em i.e.}\ $\lambda>\Gamma_S$. As
$N$ increases, the dissipative coupling increases, and we reach the
regime $\lambda<\Gamma_S$. For large atom number, we find $\lambda$
approaches $\Gamma_C$. The $\Gamma_C$ limit arises from quantum
fluctuations associated with the collective pseudospin decay through
the cavity.

There are three timescales one should consider. At short times,
quantum correlations develop as the atoms phase-lock. This can be seen
in the initial transient part of the evolution of the observables
shown in Fig.~2(b), and is characterized by the timescale
$w^{-1}$. This phase-locking time should be less than the atomic
coherence time $\Gamma_S^{-1}$ in order to observe high-visibility
fringes. There is also a long timescale provided by the collective
decay time $\Gamma_C^{-1}$. It is important to operate in the
parameter regime in which $w\gg\Gamma_S\gg\Gamma_C$.

A valid question to consider is: Why does the large incoherent
repumping rate $w$ not destroy the synchronization? Somewhat
paradoxically, repumping is crucial for building up phase correlations
among atoms.  In Fig.~\ref{Fig3}, we show the effect of $w$ on the
decay rates of the Ramsey fringe visibility $\lambda$. When the
repumping rate is too small or too large we find $\lambda>\Gamma_S$,
so that the system performs worse than conventional RS. This can be
understood since an effective Kuramoto model~\cite{sync1,sync2} for
Eq.~(\ref{mastereq}) shows that population inversion of the
pseudospins is a necessary condition for phase
synchronization~\cite{suppl}. The repumping strength must be large
enough that there is more probability for the atoms to be in the
excited state than in the ground state. However, if the repumping rate
is too large, the associated quantum noise destroys the phase
correlations before they can develop. As has also been seen in the
case of the superradiant laser~\cite{Meiser09,Thompson12}, the most
coherent system is realized at an intermediate pump strength.

An accurate semiclassical approximation may be developed that is valid
in the case of large numbers of atoms. Taking advantage of the fact
that all expectation values are symmetric with respect to atom
exchange, we find from Eq.~(\ref{mastereq}),
\begin{equation}\label{sc}
  \frac{d}{dt}\langle\hat{\sigma}_j^+\rangle
  =i\Delta\nu\langle\hat{\sigma}_j^+\rangle
  -\frac{\Gamma_t}{2}\langle\hat{\sigma}_j^+\rangle
  +\frac{\Gamma_C}{2}(N-1)\langle\hat{\sigma}_j^+\hat{\sigma}_k^z\rangle,
\end{equation}
where $j\ne k$ and $\Gamma_t=2\Gamma_S+w+\Gamma_C$ is the total decay
rate of the atomic coherence. We first point out that instead of
calculating the population difference measured at the end of the RS
sequence, it is equivalent to calculate
$2\mathrm{Im}[\langle\hat{\sigma}_j^+\rangle]$ just before the second
$\pi/2$ pulse. The decay rate of $\langle\hat{\sigma}_j^+\rangle$
during the interrogation time $T$ is therefore the same as that of the
Ramsey fringe visibility. As seen in Fig.~\ref{Fig2}(b), the
quantities $\alpha(t)=\langle\hat{\sigma}_j^+\hat{\sigma}_k^z\rangle/
(\langle\hat{\sigma}_j^+\rangle\langle\hat{\sigma}_k^z\rangle)$ and
$\langle\hat{\sigma}_j^z(t)\rangle$ rapidly approach steady state on
the short timescale of the phase-locking, $w^{-1}$. We therefore
substitute the steady-state values $\alpha_{\rm ss}$ and
$\langle\hat{\sigma}_j^z\rangle_{\rm ss}$ into Eq.~(\ref{sc}). This
produces the exponentially decaying sine function solution noted
earlier with decay constant
\begin{equation}
  \lambda=\frac12\left[\Gamma_t-(N-1)\Gamma_C
  \alpha_{\rm ss}\langle\hat{\sigma}_j^z\rangle_{\rm
    ss}\right].
\end{equation}
Furthermore $\alpha_{\rm ss}\approx1$, see Fig.~\ref{Fig2}(b). At the
level of mean-field~\cite{suppl}, $\langle\hat{\sigma}_j^z\rangle_{\rm
  ss}\approx\Gamma_t/(N-1)\Gamma_c$ giving the trivial result
$\lambda=0$. It is therefore necessary to develop a semiclassical
expression for $\langle\hat{\sigma}_j^z\rangle_{\rm ss}$ that goes
beyond mean-field, as shown in~\cite{suppl}.  Fig.~\ref{Fig3} compares
$\lambda$ from the semiclassical expression with the quantum master
equation solution, showing good agreement over the full range of
pumping rates.

All of these results consider the ensemble that is formed from a
statistical average of independent trials. The decay of the fringe
visibility is really due to the averaging itself, as we will now see. In
each trial, the quantum phase is diffusing as a function of
interrogation time. This means that as time goes on,
different trials begin to  add out of phase, and so the fringe visibility
decays.

This motivates us to consider the properties of a single experimental
run, where the behavior is qualitatively different. Although in a
single run, the fringe undergoes a quantum phase diffusion, it does so
with non-decaying visibility. This quantum phase diffusion has a
simple physical interpretation in terms of quantum measurements. Since
the cavity field is slaved to the atomic coherence through adiabatic
elimination, measuring the phase of the cavity output field, for
example by homodyne measurement, is equivalent to a continuous
non-destructive measurement on which information is gathered about the
evolving collective atomic phase. The back-action of this measurement
introduces fluctuations that cause the collective atomic phase to
undergo a random-walk~\cite{thompson13}.

\begin{figure}[t]
  \centerline{\includegraphics[width=0.7\linewidth, angle=0]{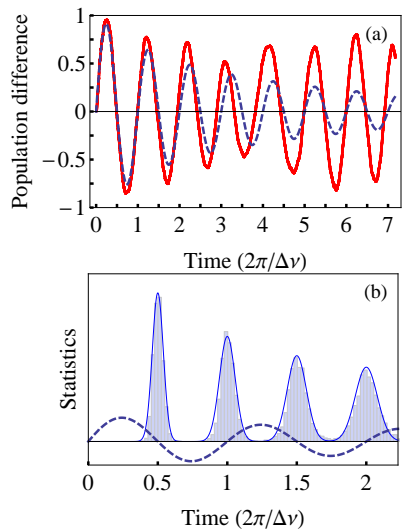}}
  \caption{\label{Fig4}(color online) Quantum state diffusion
    calculations of conditional Ramsey fringes subject to continuous
    homodyne measurement of the cavity output field for $N=10$ and
    $w=N\Gamma_C/2$. The blue dashed lines are the ensemble average
    for reference. (a) A typical Ramsey fringe for a single
    experimental trial (red solid line). (b) Histograms are the
    statistics of the positions of zero crossings of each fringe for
    4000 trials. The blue solid lines are fitted Gaussian
    distributions with variance of $T\Gamma_C$ centered on the zero
    crossing of the ensemble average.}
\end{figure}

We demonstrate this in Fig.~\ref{Fig4}(a), where we show a typical
Ramsey fringe for a single experimental trial by using the method of
quantum state diffusion~\cite{Milburn931,Milburn932} to yield
conditional evolution of the system subject to continuous measurements
of the cavity field.  The phase diffusion of the synchronized atoms is
evident from the phase fluctuation of the Ramsey fringe. To find the
phase diffusion coefficient, Fig.~\ref{Fig4}(b) shows the statistics
of the positions of the zero crossings of the fringe for 4000
trials. They fit well to Gaussian distributions with variance given by
$T\Gamma_C$, clearly demonstrating that it is a diffusion process and
that the diffusion coefficient is $\sqrt{\Gamma_C}$.  Note that this
is the same mechanism that also sets the quantum limited linewidth in
a superradiant laser to be $\Gamma_C$~\cite{Meiser09}, observed here
in the time rather than frequency domain.

We should emphasize that the quantum phase diffusion does not itself
provide a fundamental limit to the performance of conditional RS,
since the collective atomic phase can be tracked by measuring the
light output from the cavity. This opens up the exciting possibility
of observing conditional Ramsey fringes (meaning an experimental trial
conditioned on the measurement record of the output field) of near
maximum fringe visibility for as long as the atoms can be stored, even
in the presence of $T_1$ and $T_2$ processes. Of course a practical
limit is also set by the length of time for which the local oscillator
can remain phase coherent. In principle, if experimentally achieved,
this work could lead to dramatic advances in the sensitivity of RS,
since the entire measurement interval could then be used to determine
frequency at the Fourier limit.

In conclusion, we have proposed and analyzed RS with synchronized
atoms where we have shown that the interrogation time can be extended
beyond the $T_1$ and $T_2$ times that limit conventional RS.  Due to
the rephasing effect, we have demonstrated that synchronized atoms are
potentially robust against local decoherence. However, we have also
found that the synchronization process itself intrinsically generates
quantum phase diffusion through the quantum fluctuations that arise
due to the cavity dissipation. This implies that the quantum phase of
the atomic ensemble relative to the local oscillator must be tracked
in real time by observation of the output light from the cavity in
order to achieve the optimal precision for the RS with synchronized
atoms.

We acknowledge helpful discussions with J. Cooper, D. Meiser, B. Zhu,
D. A. Tieri, C. Genes, J. G. Restrepo, A. M. Rey, J. K. Thompson and
J. Ye. This work has been supported by the DARPA QuASAR program, the
NSF, and NIST.

\renewcommand*{\citenumfont}[1]{S#1}
\renewcommand*{\bibnumfmt}[1]{[S#1]}
\renewcommand{\thesection}{S.\arabic{section}}
\renewcommand{\thesubsection}{\thesection.\arabic{subsection}}
\makeatletter 
\def\tagform@#1{\maketag@@@{(S\ignorespaces#1\unskip\@@italiccorr)}}
\makeatother
\makeatletter
\makeatletter \renewcommand{\fnum@figure}
{\figurename~S\thefigure}
\makeatother

\onecolumngrid
\begin{center}
\textbf{SUPPLEMENTARY MATERIALS}
\end{center}
\setcounter{equation}{0} \emph{Effective Kuramoto model~-~}We first
make the mean-field ansatz that the density matrix is a product of
density matrices for each atom, {\it i.e.}, $\rho=\prod_j\rho_j$.  We
have checked that this ansatz is accurate to $O(1/N)$. Plugging this
ansatz into Eq.~(3) in the paper, we obtain the equation of motion for
$j$-th atom by tracing out all other atoms;
\begin{eqnarray}\label{mf}
  \frac{d\rho_j}{dt}&=&
  \frac{1}{i}[\frac{\Delta\nu}{2}\hat{\sigma}_j^{z},\rho_j]
  +\sum_{j=1}^N\Bigl(w\mathcal{L}[\hat{\sigma}_{j}^+]
  +(\frac{1}{T_1}+\Gamma_C)\mathcal{L}[\hat{\sigma}_{j}^-]
  +\frac{1}{4T_2}
  \mathcal{L}[\hat{\sigma}_{j}^z]\Bigr)\rho_j
  \nonumber\\ &&{}
  +\frac{\Gamma_C}{2}(\hat{\sigma}_j^-\rho_j
  -\rho_j\hat{\sigma}_j^-)\mathcal{O}
  +\frac{\Gamma_C}{2}(\rho_j\hat{\sigma}_j^+-\hat{\sigma}_j^+\rho_j)
  \mathcal{O}^*,
\end{eqnarray}
where $\mathcal{O}=\sum_{m\ne
  j}\langle\sigma_m^+\rangle$. Eq.~(\ref{mf}) is self-consistent since
the effect of all the other atoms is approximated by a mean field
$\mathcal{O}$.  $\mathcal{O}$ acts as an order parameter for the
synchronization phase transition: in the absence of synchronization,
or phase correlation between atoms, $|\mathcal{O}|=0$, while
$|\mathcal{O}|>0$ in the synchronized phase, breaking the $U$(1)
symmetry of Eq.~(3) in the paper.

There are two factors at work in Eq.~(S\ref{mf}), the interaction with
the mean field~[resulting from the dissipative coupling in Eq.~(3) of
the paper] and quantum noises on individual atoms~\cite{s1}. We can
see this from the quantum Langevin equation for $\hat{\sigma}_j^+$;
\begin{equation}
  \frac{d}{dt}\hat{\sigma}_j^+=i\Delta\nu\hat{\sigma}_j^+
  -\frac{1/T_1+1/T_2+w+\Gamma_C}{2}\hat{\sigma}_j^++
  \frac{\Gamma_C}{2}\mathcal{O}\hat{\sigma}_j^z+\mathcal{F}(t),
\end{equation}
where $\mathcal{F}(t)$ is the quantum noise contributed by spontaneous
emission, inhomogeneous dephasing, repumping and collective decay. The
quantum noises randomize the phase of individual atoms, and thus
inhibit phase locking between atoms. To find the effect of the
dissipative coupling between atoms, we parameterize
$\langle\hat{\sigma}_j^+\rangle$ as $\alpha_je^{-i\phi_j}$ and derive
the equation of motion for $\phi_j$,
\begin{equation}\label{kura}
  \frac{d}{dt}\phi_j=-\Delta\nu+\frac{\Gamma_C}{2}
  \frac{\langle\hat{\sigma}_j^z\rangle}{\alpha_j}
  \sum_m{\alpha_m}\sin(\phi_m-\phi_j).
\end{equation}
Eq.~(S\ref{kura}) is equivalent to the well-known Kuramoto
model~\cite{s2} for describing the phase synchronization.  In the case
of $\langle\hat{\sigma}_j^z\rangle>0$, the coupling gives rise to
phase attraction between atoms.

\vspace{0.5cm}
\noindent\emph{Semiclassical approximation for 
$\langle\hat{\sigma}_j^z\rangle_{\mathrm{ss}}$~-~}To find
$\langle\hat{\sigma}_j^z\rangle_{\mathrm{ss}}$, we employ the cumulant
approximation method~\cite{s3}. Expectation values of the atoms are
expanded in terms of $\langle\hat{\sigma}_j^z\rangle$ and
$\langle\hat{\sigma}_j^+\hat{\sigma}_k^-\rangle$. Their equation of
motion can then be found from Eq.~(3) in the paper,
\begin{eqnarray}
  \frac{d}{dt}\langle\hat{\sigma}_{j}^z\rangle= 
  &&-(\Gamma_C+\frac{1}{T_1})\left(\langle\hat{\sigma}_j^z\rangle+1\right)
  -w\left(\langle\hat{\sigma}_j^z\rangle-1\right)
  -2\Gamma_C(N-1)\langle\hat{\sigma}_j^+\hat{\sigma}_k^-\rangle \nonumber\\
  \frac{d}{dt}\langle\hat{\sigma}_j^+\hat{\sigma}_k^-\rangle\approx 
  &&-\Gamma_t\langle\hat{\sigma}_j^+\hat{\sigma}_k^-\rangle+
  \frac{\Gamma_C}{2}\langle\hat{\sigma}_j^z\rangle
  \left(1+\langle\hat{\sigma}_j^z\rangle\right)
  +\Gamma_C(N-2)\langle\hat{\sigma}_j^+\hat{\sigma}_k^-\rangle
  \langle \hat{\sigma}_j^z\rangle.\nonumber
\end{eqnarray}
where we have factorized
$\langle\hat{\sigma}_j^z\hat{\sigma}_k^z\rangle\approx
\langle\hat{\sigma}_j^z\rangle^2$
and
$\langle\hat{\sigma}_j^+\hat{\sigma}_k^-\hat{\sigma}_l^z\rangle\approx
\langle\hat{\sigma}_j^+\hat{\sigma}_k^-\rangle\langle
\hat{\sigma}_j^z\rangle$.
$\langle\hat{\sigma}_j^z\rangle_{\mathrm{ss}}$ can then be found by
setting the time derivatives to zero, and the resulting algebraic
equations form a close set and can be solved exactly.


\begin{thebibliography}{PRL}
\bibitem{s1}Note that the frequency disorder of atoms is absent from
  the current model. In future work, it will be interesting to
  investigate the dependence of synchronization on the frequency
  disorder.
\end{thebibliography}
\end{document}